\documentclass[review]{elsarticle}

\usepackage{lineno,hyperref}
\usepackage{amsfonts}
\usepackage{amsmath}
\usepackage{amssymb}
\usepackage{mathtools}
\usepackage{feynmp}
\usepackage{feynmp-auto}
\usepackage{graphicx}
\usepackage{wrapfig}
\usepackage{color}
\usepackage[title,titletoc,toc]{appendix}

\usepackage[utf8]{inputenc}
\usepackage[english]{babel}
\usepackage[euler]{textgreek}

\usepackage{multicol}
\modulolinenumbers[5]

\journal{Physics Letters B}

\begin{document}

\begin{frontmatter}

\title{Improved constraints on monopole-dipole interaction mediated by pseudo-scalar bosons}

\author{N. Crescini\fnref{email}}
\fntext[email]{nicolo.crescini@phd.unipd.it}
\address{
 Dipartimento di Fisica e Astronomia, Via Marzolo 8, I-35131 Padova (Italy)\\  and INFN, Laboratori Nazionali di Legnaro, Viale dell'Universit\`a 2, I-35020 Legnaro, Padova (Italy)}

\author{C. Braggio}
\author{G. Carugno}
\address{INFN, Sezione di Padova and Dipartimento di Fisica e Astronomia, Via Marzolo 8, I-35131 Padova (Italy)}

\author{P. Falferi}
\address{Istituto di Fotonica e Nanotecnologie, CNR—Fondazione Bruno Kessler, and
INFN-TIFPA, I-38123 Povo, Trento (Italy)}

\author{A. Ortolan}
\author{G. Ruoso}
\address{INFN, Laboratori Nazionali di Legnaro, Viale dell'Universit\`a 2, I-35020 Legnaro, Padova (Italy)}

\begin{abstract}
We present a more stringent upper limit on long-range axion-mediated forces obtained by the QUAX-g$_p$g$_s$ experiment, located at the INFN - Laboratori Nazionali di Legnaro. By measuring variations of a paramagnetic GSO crystal magnetization with a dc-SQUID magnetometer we investigate the possible coupling between electron spins and unpolarized nucleons in lead disks. The induced magnetization can be interpreted as the effect of a long-range spin dependent interaction mediated by axions or Axion Like Particles (ALPs). The corresponding coupling strength is proportional  to the CP violating term $g_p^eg_s^N$, i.e. the product of the pseudoscalar and scalar coupling constants of electron and nucleon, respectively. Our upper limit is more constraining than previous ones in the interaction range $0.01\,\mathrm{m}<\lambda_a<0.2\,$m, with a best result on $g_p^eg_s^N/(\hbar c)$ of $4.3\times10^{-30}$ at 95\% confidence level in the interval $0.1\,\mathrm{m}<\lambda_a<0.2\,$m. We eventually discuss our  plans to improve the  QUAX-g$_p$g$_s$ sensitivity by a few orders of magnitude, which will allow us to investigate the $\vartheta\simeq 10^{-10}$ range of CP-violating parameter and test some QCD axion models.
\end{abstract}


\end{frontmatter}

\linenumbers

\section{Introduction}
\label{intro}
\begin{figure}[h]
\begin{center}
\begin{fmffile}{gpgs}
\begin{fmfgraph*}(120,60)
\fmfleft{i1,i2}
\fmfright{o2,o1}
\fmflabel{$e^-$}{i1}
\fmflabel{$e^-$}{i2}
\fmflabel{$N$}{o1}
\fmflabel{$N$}{o2}
\fmflabel{$ig_p^e\gamma_5$}{v1}
\fmflabel{$ig_s^N$}{v2}
\fmf{fermion}{i1,v1,i2}
\fmf{fermion}{o2,v2,o1}
\fmf{dashes,label=$a$}{v1,v2}
\end{fmfgraph*}
\end{fmffile}
\end{center}
\caption{\footnotesize Interaction  diagram  of a scalar-pseudoscalar coupling between a nucleus $N$ and an electron $e^-$. $N$ is unpolarized and interacts at the scalar vertex with the coupling constant $g^N_s$, whereas $e^-$ is polarized and interacts at the pseudoscalar vertex with the coupling constant $g^e_p$. Here the mediator is the axion $a$ and the interaction strength is proportional to $g^N_s g^e_p$.}
\label{fig1}
\end{figure}
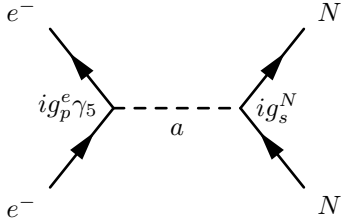

The signature of  symmetry breaking at extremely high energies can be highlighted by the presence of long-range ultraweak forces mediated by pseudo-Goldstone bosons \cite{weinberg}. In particular, the pseudo-boson can be either the QCD axion or an axion-like-particle (ALP), which involves P and T violating forces with strength proportional to the product of the couplings at the pseudo-boson vertices \cite{wilczek}. There are two options for coupling pseudo-scalar bosons with fundamental fermions: i) the  spin-dependent pseudoscalar vertex, and ii) the scalar vertex that becomes spin-independent in the non relativistic limit. Thus, in a multipole expansion, the two fields are described by the ``dipole'' (pseudo-scalar coupling $g_p$) and ``monopole'' (scalar coupling $g_s$) moments, respectively. For instance, exchange of virtual axions - a possible solution of the strong CP problem - mediates a monopole-dipole force where $g_s$ is proportional to the QCD vacuum angle $\vartheta\simeq 10^{-10} \div 10^{-14}$.
In Fig.(\ref{fig1}) we report the Feynman diagram of the $g_p g_s$ interaction between an electron $e^-$ and a nucleus $N$ mediated by an axion or ALPs that we investigate in this paper.
\noindent The monopole-dipole coupling of the spin of a polarized electron with an unpolarized nucleus mediated by axions is described by the potential \cite{wilczek}
\begin{equation}
V_{md}(\boldsymbol{r})=\frac{\hbar g_p^e g_s^N}{8 \pi m_e c} \Big[ ( \boldsymbol{\hat{\sigma}} \cdot \boldsymbol{\hat{r}}) \Big(\frac{1}{r\lambda_a}+\frac{1}{r^2} \Big) \Big] e^{-r/\lambda_a},
\label{Vmd}
\end{equation}
where $\lambda_a$ is the Compton wavelength of the axion (interaction range), $g_p^e$ and $g_s^N$ are the coupling constants of the interaction, $\hbar$ is the reduced Plank constant, $c$ is the speed of light in vacuum, $m_e$ is the mass of the electron, $ \boldsymbol{\hat{\sigma}}$ is the vector of Pauli spin matrices, and $r$ and $\boldsymbol{\hat{r}}$ are the distance and unit vector between the nucleon and the electron, respectively \cite{weinberg,wilczek,PhysRevD.52.3132}. 
It is worth noticing that for the axion the expected values of $g_p^e g_s^N$ coupling strength as a function of its mass $m_a$ is \cite{wilczek}
\begin{equation}
g_p^e g_s^N=\frac{\vartheta \sigma}{f_a^2} \frac{m_u m_d}{(m_u+m_d)^2} m_a ,
\label{gpgs_int}
\end{equation}
where $m_u$ and $m_d$ are the masses of the up and down quarks. In the conservative Kobayashi-Maskawa model, 
the predicted value of the vacuum angle is $\vartheta\sim10^{-14}$, the pion-nucleon $\sigma$ term is taken to be $60\,$MeV \cite{Sainio1995}, $f_a$ is the breaking energy scale of the Peccei-Quinn symmetry, and so the coupling strength should be $g_p^e g_s^N/(\hbar c)\sim 10^{-37} (m_a/ 1\,\mu$eV).
If the mediator mass is sufficiently small ($m_a\lesssim 10^{-5}$eV),  $\lambda_a = h/m_a c \gtrsim 0.1\,$m is  macroscopic  and a long-range force arises. Many attempts have been done to measure this spin-matter force in laboratory experiments over recent decades. 
However, only a few experiments for length scales $0.1\div20$\,cm have placed upper limits on the product coupling between systems of electron spins and  unpolarized nuclei by exploiting different approaches, for instance dc-SQUIDs and paramagnetic salts \cite{ni}, high precision torsion balances \cite{PhysRevLett.106.041801,PhysRevLett.98.081101,PhysRevLett.70.701,Daniels1994149}, atomic magnetometers \cite{PhysRevLett.77.2170,PhysRevLett.68.135} and stored ion spectroscopy \cite{PhysRevLett.67.1735}.
New experiments have also been proposed \cite{PhysRevLett.113.161801,PhysRevD.91.102006} that should be able to reach better sensitivities.

\begin{figure}[h!]
\centering
\includegraphics[width=.6\textwidth]{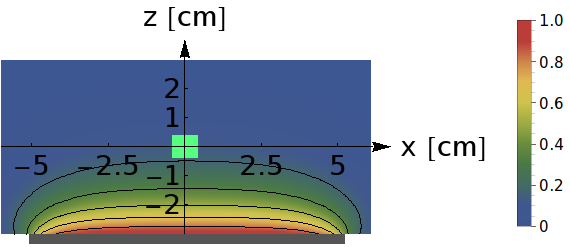}
\caption{\footnotesize Relative amplitude of the z-component of the effective magnetic field $B_\mathrm{eff}$ produced by a disk of diameter $D=10$\,cm, thickness $d=2.5$\,cm (dark gray rectangle) assuming $\lambda_a=1$\,cm. The integral defining $B_\mathrm{eff}$ has been numerically evaluated in Mathematica 10. The paramagnetic crystal is represented as a green square located at the origin of the coordinate system.}
\label{source}
\end{figure}

Eq.(\ref{Vmd}) can be conveniently recast as the energy of electron magnetic moment  $\boldsymbol{\mu}\equiv \mu_B \boldsymbol{\hat{\sigma}}$ in the effective magnetic field 
\begin{equation}
\boldsymbol{b}_{\mathrm{eff}}(\boldsymbol{r}) = - \frac{g_p^e g_s^N}{4 \pi e c}   \boldsymbol{\hat{r}} \Big(\frac{1}{r\lambda_a}+\frac{1}{r^2} \Big) e^{-r/\lambda_a},
\label{beff}
\end{equation}
where $\mu_B$ is Bohr's magneton and $e$ is the electron charge.  Clearly, this field is not a genuine magnetic field,  as the interaction potential is generated by pseudoscalar exchange rather than by photon exchange, and so it does not satisfy the Maxwell’s equations.
Once Eq.(\ref{beff}) is integrated over a macroscopic monopole source of volume $V_S$ with $ N\simeq O(10^{23})$ nuclei and $\rho_N$ nucleon density, the resulting total effective magnetic field
 \begin{equation}
\boldsymbol{B}_{\mathrm{eff}}(\boldsymbol{r})=-\frac{g_p^e g_s^N \rho_N}{4 \pi e c} \iiint \limits_{V_S} d x'd y'd z' \ (\boldsymbol{\hat{r}-\hat{r}'}) \Biggl[\frac{1}{|\boldsymbol{r}-\boldsymbol{r}'|\lambda_a}+\frac{1}{|\boldsymbol{r}-\boldsymbol{r}'|^2} \Biggr] e^{-|\boldsymbol{r}-\boldsymbol{r}'|/\lambda_a} \ ,
\label{beffs}
\end{equation}
can have a measurable amplitude. 
In fact, the macroscopic magnetization induced by this field on polarizable electrons of a paramagnetic material (detector), with magnetic susceptibility $\chi=O(1)$ and negligible dimensions with respect to the source volume $V_S$, reads $\mu_{_0}M= \chi B_{\mathrm{eff}}$. The component $B_\mathrm{eff}$ along the line joining the center of mass of source and detector is calculated by direct numerical integration of Eq.(\ref{beffs}) and $\mu_{_0}$ is the magnetic permeability of vacuum, as reported in Fig.(\ref{source}). We will see in Sect.(\ref{analysis_res}) that magnetizations as low as $\mu_{_0}M\simeq 10^{-17}$\,T can be measured with a dc-SQUID operated as a magnetometer with integration time of $10^4$\,s, which is enough to improve current upper limits on $g_p^eg_s^N$ of one order of magnitude.

The plan of the paper is as follows.  In Section \ref{apparatus} we describe the QUAX-g$_p$g$_s$ apparatus and  the effect of the long range, spin dependent interaction on our observable, i.e. the magnetization induced on the detector by an effective magnetic field source. In Section \ref{analysis_res} we discuss our data analysis and the results we obtained with the present experimental set-up. Conclusions and plans to improve the QUAX-g$_p$g$_s$ sensitivity are presented in Section \ref{conc}.

\section{The QUAX-g$_p$g$_s$ apparatus}
\label{apparatus}
\begin{figure}[h!]
\centering
\includegraphics[width=.4\textwidth]{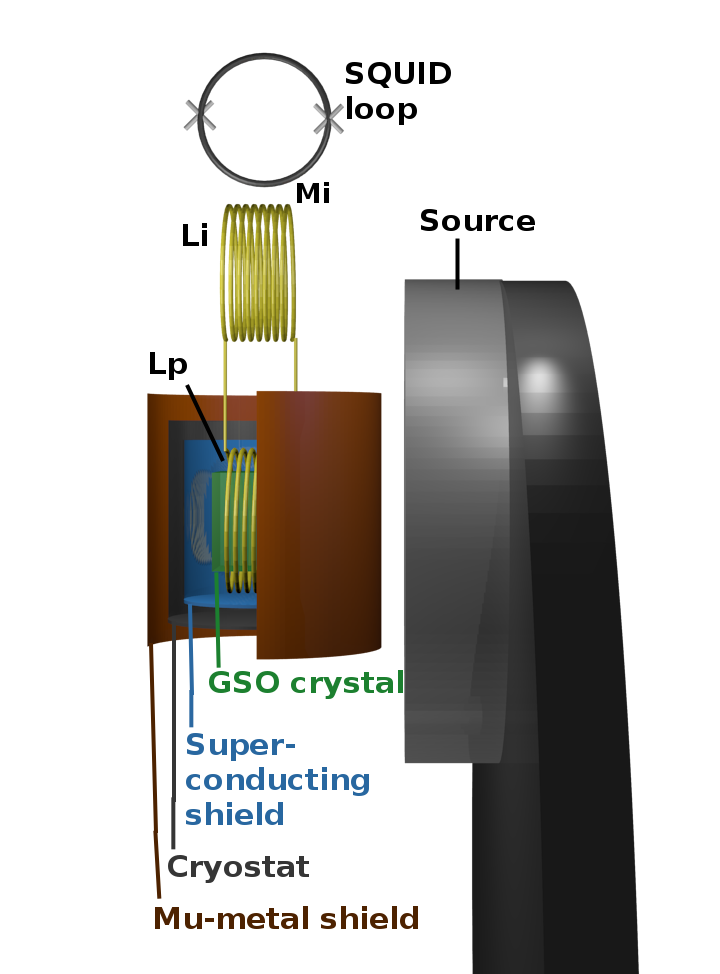}
\includegraphics[width=.55\textwidth]{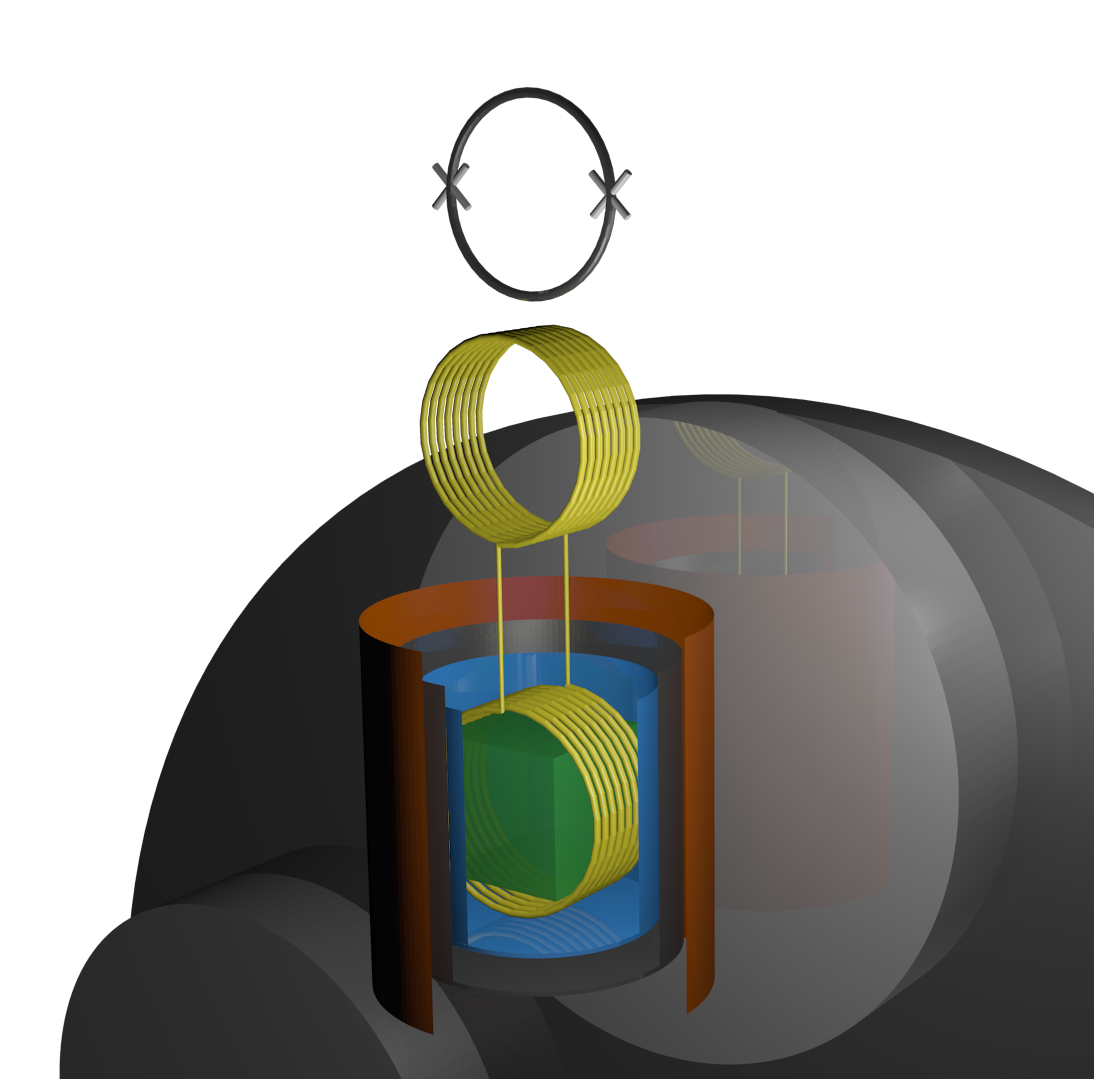}
\caption{\footnotesize Schematic model of the apparatus. $L_i=1.8\,\mu$H is the input coil of the SQUID, $L_p\simeq1.8\,\mu$H is the pick-up coil wound on the GSO crystal. On the right side of the figure is represented a source mounted on the rotating wheel.}
\label{app_yomo}
\end{figure}

The experiment (see Ref.\cite{Crescini2017109}  for more details) is performed by measuring the magnetization of a cubic sample of gadolinium oxyorthosilicate Gd$_2$SiO$_5$ crystal (GSO) with $1$\,cm edge length (detector), induced by $N_s=4$ disk shaped lead masses (sources).
GSO is a paramagnetic material with a magnetic susceptibility $\chi\simeq0.7$ \cite{Crescini2017109} at cryogenic temperatures. The crystal is housed in the lower part of a liquid helium cryostat [see Fig.(\ref{app_yomo})] and cooled down to $\simeq4\,$K. 
The distance between the center of mass of each $B_\mathrm{eff}$ source and the GSO crystal is modulated in time by mounting the masses on a rotating aluminum wheel as illustrated in  Fig.(\ref{app_yomo}).
The aluminum wheel is $70\,$cm in diameter and rotates at a constant angular velocity.
The minimum distance between each source and detector is $3.7$\,cm.

To detect the variation of magnetization we use the most sensitive magnetometer available, namely a dc-SQUID operated at  $\sim$4\,K.
As shown in Fig.(\ref{app_yomo}), the superconducting input coil of the SQUID $L_i\simeq1.8\,\mu$H is connected to a superconducting pick-up coil $L_p$, which is wound around the GSO crystal. To satisfy the optimal matching condition $L_p=L_i$ of the SQUID, the pick-up coil is made of 8 turns of a NbTi wire.
The two coils $L_i$ and $L_p$ are connected in series forming  a superconducting transformer which transfers the magnetic flux from the pick-up coil to the SQUID loop.
As $B_\mathrm{eff}$ is not a true magnetic field, we can reduce the environmental magnetic disturbances around the GSO with magnetic shields. 
In particular, we make use of two concentric Bi-2223 cylindrical superconducting shields at liquid helium temperature, and a $\mu$-metal external shield at room temperature to reduce the field trapped in the inner shields. The overall rejection factor of the two superconducting shields is expected to be $\sim 10^{12}$ \cite{magshield}, which is sufficient to make environmental magnetic disturbances negligible.

We calibrated the apparatus using a solenoid with a diameter of $5\,$cm, coaxial to the pick-up coil, providing a uniform magnetic field over $L_p$. The resulting conversion factor between the output voltage of the SQUID electronics and the magnetic field at the pick-up is $4.25\times10^{-11}\,$T/V. More details about the SQUID readout can be found in Ref.\cite{Crescini2017109}.
The SQUID output is fed to a band-pass filter having lower and upper cutoff frequencies of $0.1\,$Hz and $25\,$Hz.Measurements are taken at 10\,Hz, well above the 1/f noise knee of the SQUID \cite{magnicon}.
In our experimental setup, the dominant noise source is the additive flux noise of the SQUID, which therefore represents the sensitivity limit of the magnetometer.

\section{Data analysis and results} 
\label{analysis_res}
\begin{figure}[h!]
\centering
\includegraphics[width=.45\textwidth]{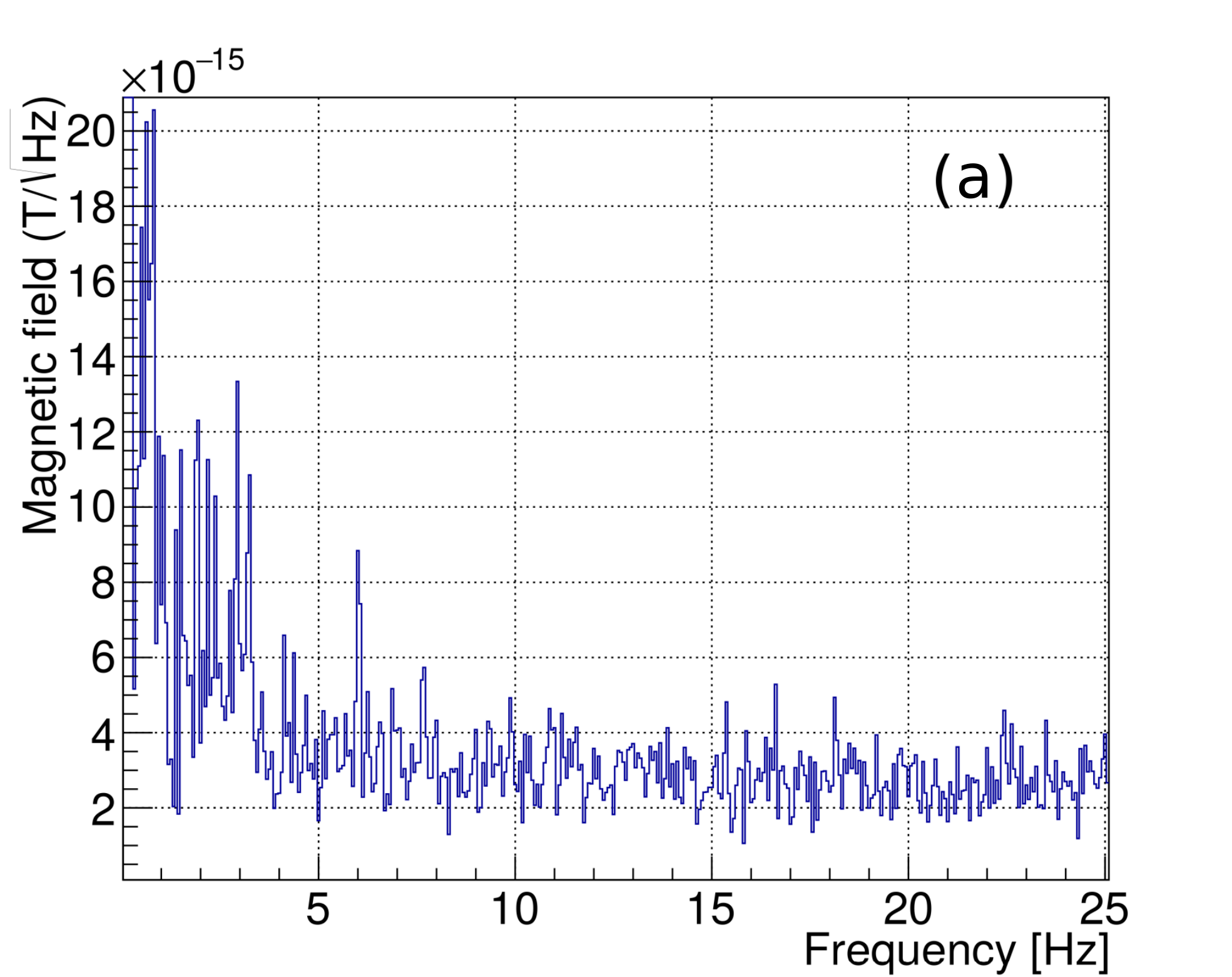} \includegraphics[width=.5\textwidth]{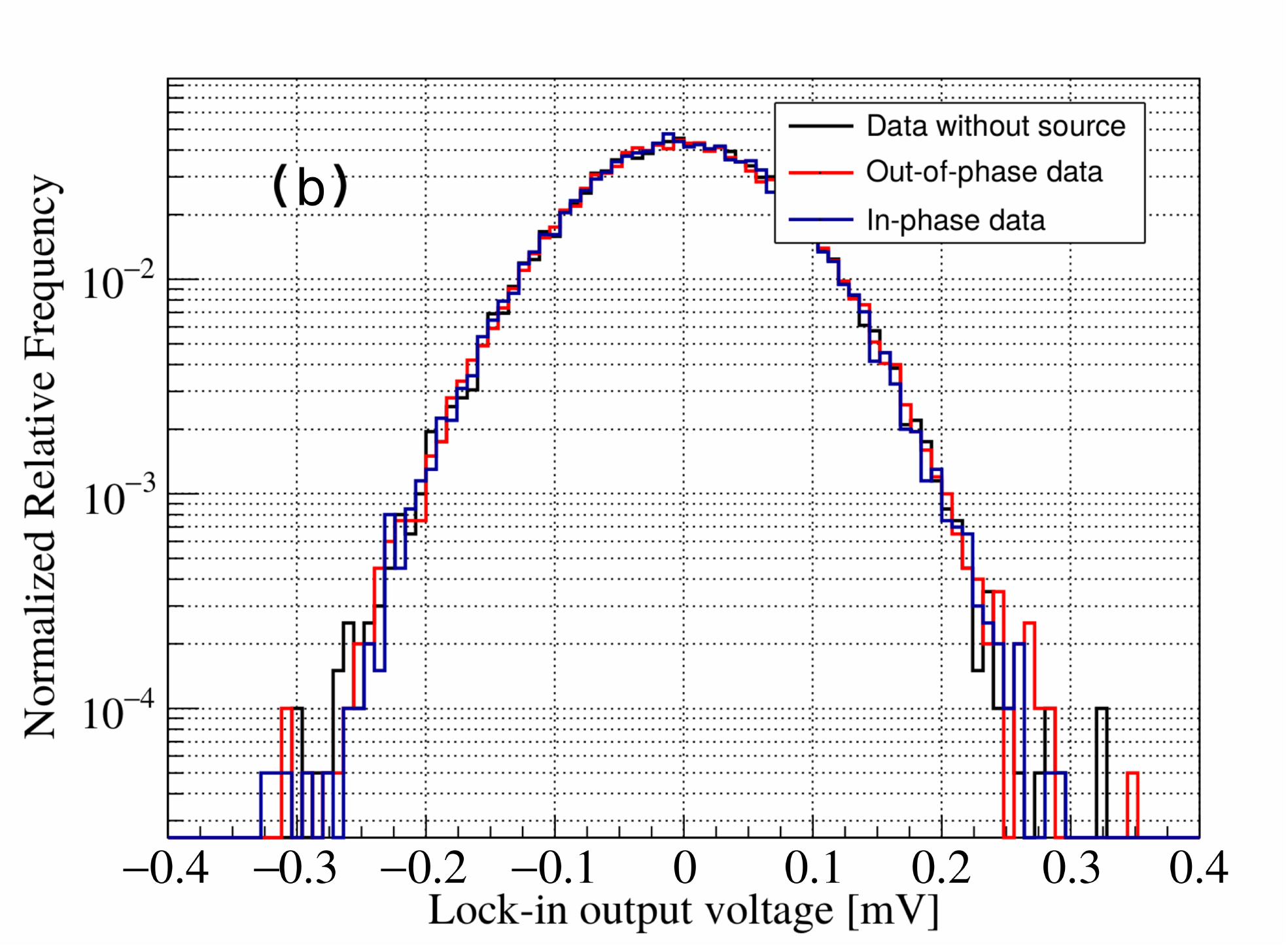}
\caption{\footnotesize
(a) Power spectrum density of the equivalent magnetic field noise at the GSO pick-up coil in the frequency band $0.1\div 25$\,Hz, estimated by averaging 5 periodograms of 8\,s. The peaks at low frequencies are probably due to mechanical resonances and does not affect the measurement.
(b) 15000 samples gaussian distribution of the digital lock-in output with respect to the rotating wheel for the in-phase (blue) and out-of-phase (red) components, using a low-pass filter cutoff of 1\,s. The black curve represents the in-phase lock-in output with the lead disks removed. The averages $\pm$ standard deviations read $(5.4\times10^{-4}\pm 6.8\times10^{-2})\,$mV (blue), $(4.8\times10^{-4} \pm 6.8\times10^{-2})\,$mV (red) and $(5.0\times10^{-4}\pm 6.8\times10^{-2})\,$mV (black).}
\label{trens_spec}
\end{figure}

In Fig.(\ref{trens_spec}) we report the measured noise, which is compatible with the additive flux noise of the SQUID. Currently this fixes the sensitivity limit of the magnetometer \cite{Crescini2017109}.
We tested the hypothesis whether the wheel rotation may introduce an excess noise by comparing measurements obtained with rotating or non rotating wheel. We found no modification of the magnetic noise level in the frequency band of the measurement.
In addition, multiple measurements were taken and no time dependence of the output has been found.

To obtain an optimal estimate of the amplitude of the modulated effective magnetic field, we performed a phase  sensitive detection with a digital lock-in and the reference phase of the rotating wheel.  
To estimate this phase, we drilled 64 holes of millimeter size evenly spaced on the circumference of the wheel.  A passing-through laser illuminates a hole and a photodiode measures the intensity of the passing-through light. From the acquired intensity, we were able to estimate the rotation frequency $f_w$ of the wheel and the phase of the sources.
\begin{figure}[h!]
\centering
\includegraphics[width=\textwidth]{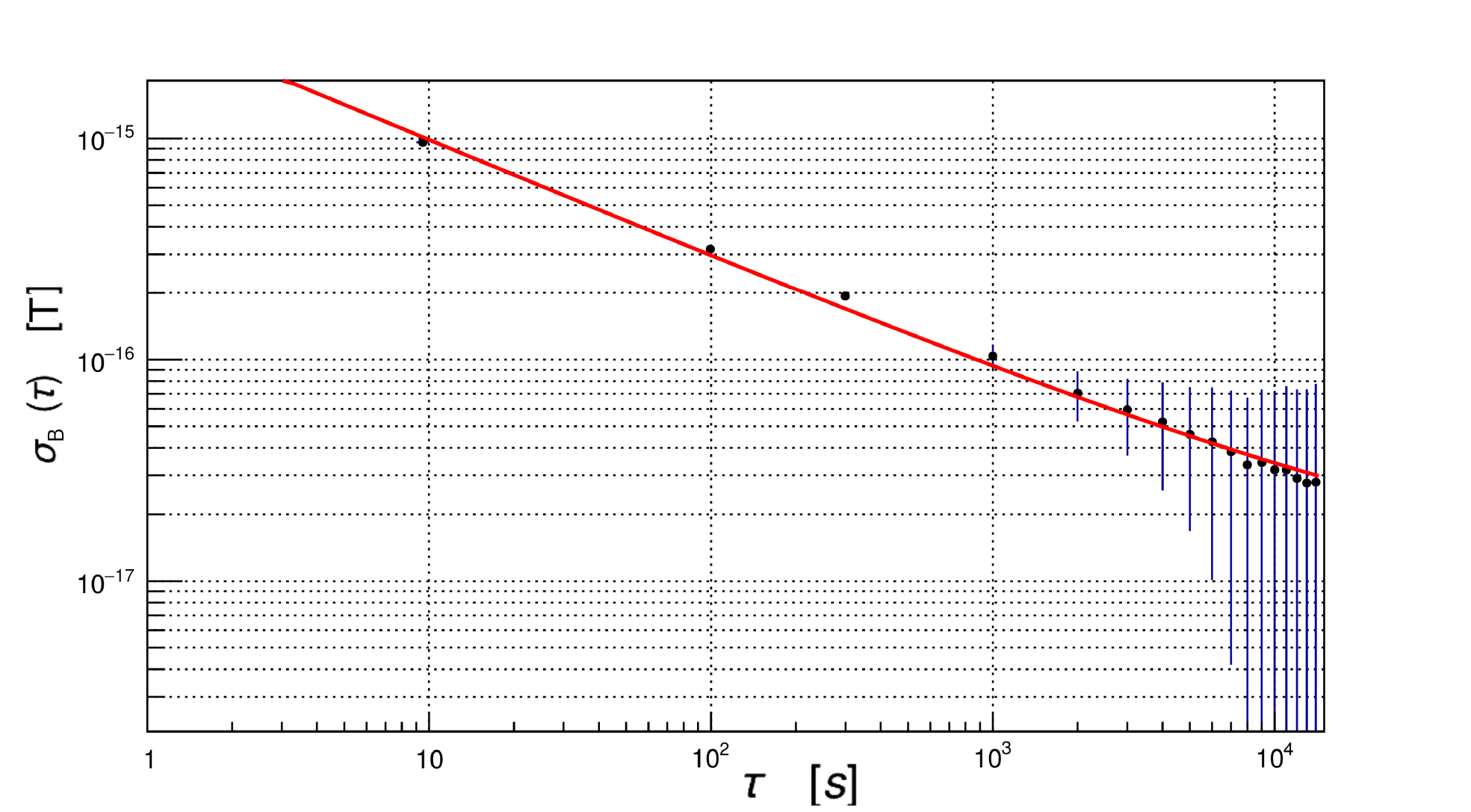}\\
\caption{\footnotesize
Allan standard deviation of the in-phase component of the digital lock-in output;  the fitting function (red) is $\sigma(\tau)=\sigma_0 \tau^{-1/2}$, showing a long term stability of the apparatus of the order of $10^4$ s (see text for further details).}
\label{allan}
\end{figure}

The SQUID output signal and the phase of the sources were acquired simultaneously with a high definition oscilloscope. Then we performed a phase sensitive detection using a digital lock-in that provides amplitude and phase of the output with respect to the signal frequency given by $f_s=4 f_w=10\,$Hz. The presence of a signal due to changes of GSO magnetization in phase with the reference can be detected looking at the statistical distribution of the lock-in output amplitudes, see Fig.(\ref{trens_spec}b). 
In Fig.(\ref{allan}) we report the Allan standard deviation of the in-phase lock-in output with reference frequency $f_s$. 
The long term stability of our apparatus allows us to integrate the lock-in output for $1.5\times10^4$\,s estimating the corresponding mean $\langle B_\mathrm{eff}\rangle= 1.8\times10^{-17}$\,T  and standard deviation $\sigma_{B_\mathrm{eff}}=2.9\times 10^{-17}$ T, expressed in equivalent magnetic field at the pickup coil. The mean value is compatible with zero within one standard deviation, and so we conclude that we have observed no induced magnetization in the GSO crystal due to monopole-dipole interaction mediated by axions or ALPs.



\subsection{Results}
Using numerical integration of Eq.(\ref{beffs}) over the volume of the sources and taking into account the geometry of the apparatus, we can convert our measurements of the effective magnetic field in a upper limit on $g_p^eg_s^N$. Since the modulation and intensity of the signal both depend on $\lambda_a$, a correction curve $g(\lambda_a)$ was estimated to obtain the actual measured limit.
Using this procedure we get our best upper limit on the coupling $g_p^eg_s^N/(\hbar c) \le 4.3\times 10^{-30}$ at 95\% C.L. in the range $1\,\mathrm{cm}<\lambda_a<20\,\mathrm{cm}$. Above the upper end of this range, the effective magnetic field no longer depends on $\lambda_a$ and so the sensitivity of our experiment to axion or ALP mediators decreases. At the lower end the sensitivity is limited by the exponential decay $e^{-r/\lambda_a}$ of the monopole dipole interaction.

\begin{figure}[h!]
\centering
\includegraphics[width=\textwidth]{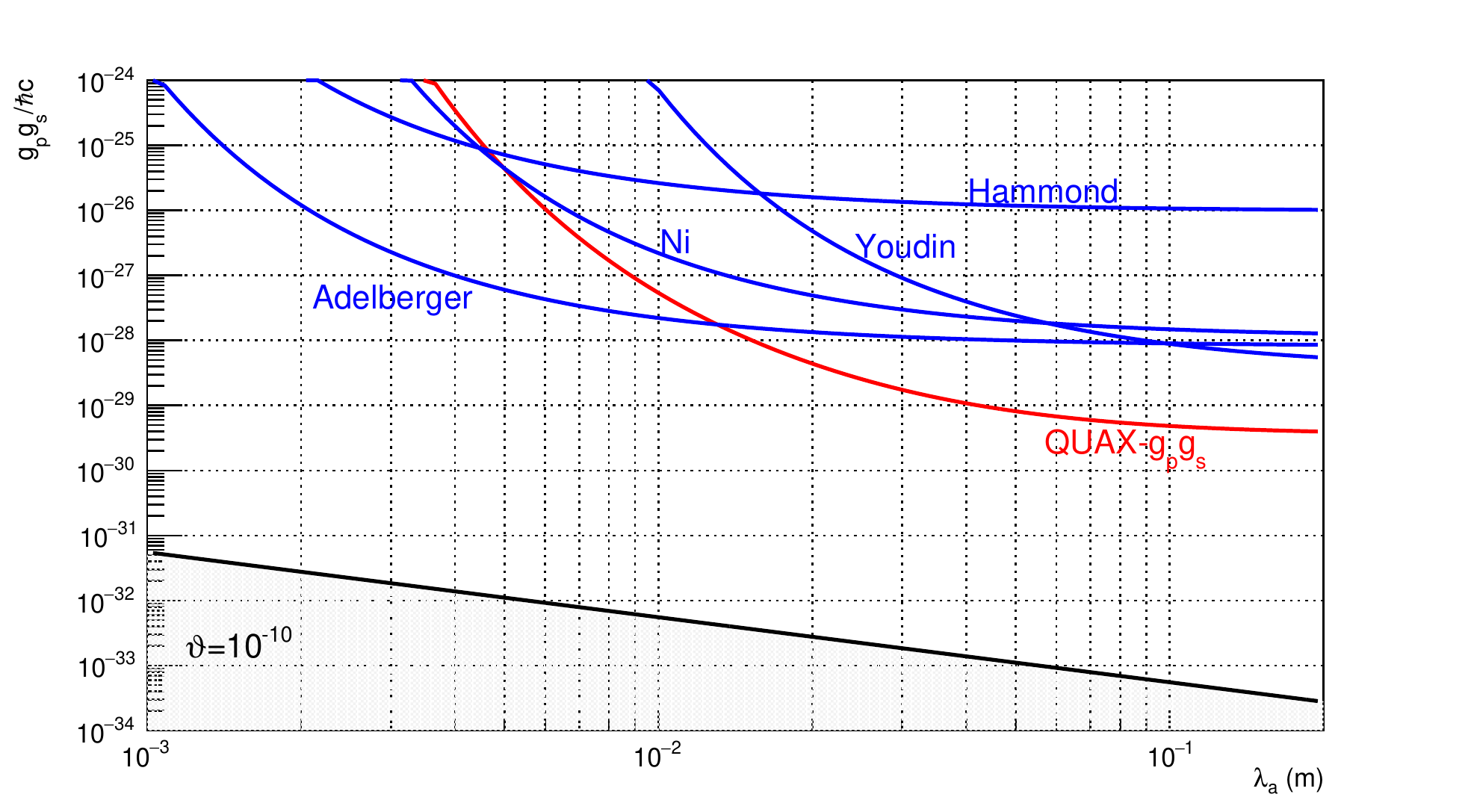}
\caption{\footnotesize Exclusion plots of monopole-dipole coupling vs. $\lambda_a$. The upper limit on the $g_p^eg_s^N$ coupling is lowered of more than one order of magnitude in respect to the previous measurements for $\lambda_a\sim10\,$cm (red line).
We also show in transparent grey the $g^e_pg^N_s$ limit derived from Eq.(\ref{gpgs_int}) with $\vartheta\le10^{-10}$ resulting from neutron EDM \cite{Baker:2006ts,PhysRevD.92.092003}, and upper limits already reported in the literature: Hammond \cite{PhysRevLett.98.081101}, Youdin \cite{PhysRevLett.77.2170}, Ni \cite{PhysRevLett.82.2439} and Adelberger \cite{PhysRevLett.115.201801}.
}
\label{sens_q}
\end{figure}

Fig.(\ref{sens_q}) compares our result to upper limits from other experiments reported in the literature
\cite{PhysRevLett.98.081101,PhysRevLett.82.2439,PhysRevLett.77.2170,PhysRevLett.70.701,PhysRevLett.68.135,PhysRevLett.115.201801} in terms of the strength of the monopole-dipole interaction.
It is worth noticing that the $g_p^eg_s^N$ coupling can be also strongly constrained by star cooling processes \cite{PhysRevD.86.015001}.

\section{Conclusions and perspectives}
\label{conc}
We have reported on a measurement of the $g^e_p g^N_s$ coupling obtained with the QUAX-g$_p$g$_s$ experiment that places an upper limit for this type of interaction. Currently, our limit is the best for spin-dependent forces mediated by axions or ALPs, as we obtained an enhancement of magnetic field sensitivity of one order of magnitude with respect to other experiments reported in the literature.
Despite the experimental approach in this paper is similar to that of Ref.\cite{ni}, we succeeded in improving the sensitivity by using  a SQUID with an intrinsic lower noise and a paramagnetic crystal with a higher susceptibility.

The sensitivity of the QUAX-g$_p$g$_s$ experiment can be further improved by using a resonant electrical LC circuit with a high quality factor $Q$ \cite{paolof2}.
In fact, we can add a coil of inductance $L\simeq100\,$mH wound around the GSO and connected to a low loss capacitor with $C\simeq 20\,\mu$F, to form an LC circuit with resonance frequency $f_\mathrm{LC}\simeq 110\,$Hz.
The pickup coil $L_p$ is also wound on the crystal and connected to the SQUID input coil $L_i$ as described in this paper.
The coupling with the SQUID reduces the value of $L$ and, when $L$ and $L_p$ are perfectly coupled, the resonance frequency increases up to the maximum value $\sqrt{2}f_\mathrm{LC}\sim$160\,Hz.
By increasing $f_w$ to 6.5\,Hz and $N_s$ to 24, the signal frequency will coincide with the LC resonance frequency. In this case, the SNR at the SQUID input coil could increase up to a factor of $Q$ with respect to the present configuration.
Such sensitivity improvement is effective as long as the Johnson and crystal magnetization noises exceed the additive noise of the SQUID.


Our improved experimental apparatus could eventually invade the gray exclusion region in the g$_p$g$_s-\lambda_a$ plane of Fig.(\ref{sens_q}), which has been established by the limit of the anomalous neutron EDM $d_n \le 3\times10^{-26}\,e\,$cm \cite{Baker:2006ts,PhysRevD.92.092003}.
As a final remark, we mention that QUAX-g$_p$g$_s$ can also explore the dipole-dipole coupling $g_p^eg_p^N$. To this aim, it is sufficient to replace the unpolarized sources of effective magnetic field mounted on the wheel with spin-polarized sources. Conceptually new designs for spin-polarized masses based on permanent magnets are reported in the literature \cite{speake1}. These and other related issues will be the subject of a forthcoming paper.
\section*{Acknowledgments}
It is a pleasure to thank Mario Tessaro, Fulvio Calaon, Marco Romanato, and Enrico Berto for their technical support in the design and construction of the apparatus. We also thank Clive Speake and Wei-Tou Ni for stimulating discussions on theoretical and experimental aspects concerning the measure of monopole-dipole coupling.

\section*{References}
\bibliographystyle{model1-num-names}
\bibliography{quax_gpgs_fr}

\end{document}